\documentclass{article}
\usepackage[accepted]{vietnam} 
\usepackage{natbib}
\usepackage{graphicx}      
\usepackage{wrapfig}
\def\farcs{\hbox{$.\!\!^{\prime\prime}$}}

\def\gtrsim{\mathrel{\hbox{\rlap{\hbox{\lower4pt\hbox{$\sim$}}}\hbox{$>$}}}}
\def\lessim{\mathrel{\hbox{\rlap{\hbox{\lower4pt\hbox{$\sim$}}}\hbox{$<$}}}}
\def\Journal{}

\begin{document}
\twocolumn[
\title{SMA and ALMA Studies of Disk- and Planet Formation around Low-mass Protostars}

\author{Shigehisa Takakuwa$^{1,2}$, Hsi-Wei Yen$^{2,3}$, Ti-Lin Chou$^{2,4}$,
Nagayoshi Ohashi$^{2,5}$, Yusuke Aso$^{5,6}$, Patrick M. Koch$^{2}$,
Ruben Krasnopolsky$^{2}$, Paul T. P. Ho$^{2,7}$, Hauyu Baobab Liu$^{2,3}$,
Naomi Hirano$^{2}$, Pin-Gao Gu$^{2}$, Chin-Fei Lee$^{2}$,
Evaria Puspitaningrum$^{8}$, Yuri Aikawa$^{9}$, Masahiro N. Machida$^{10}$,
Kazuya Saigo$^{11}$, Masao Saito$^{12}$, Kengo Tomida$^{13}$,
\& Kohji Tomisaka$^{14}$}{takakuwa@sci.kagoshima-u.ac.jp}

\address{1. Department of Physics and Astronomy, Graduate School of Science and Engineering,
Kagoshima University, 1-21-35 Korimoto, Kagoshima, Kagoshima 890-0065, Japan;
takakuwa@sci.kagoshima-u.ac.jp;
2. Academia Sinica Institute of Astronomy and Astrophysics, P.O. Box 23-141, Taipei 10617, Taiwan;
3. European Southern Observatory, Karl-Schwarzschild-Str. 2, Garching 85748, Germany;
4. Kavli Institute for Cosmological Physics and The Enrico Fermi Institute, The University of Chicago, Chicago, IL, USA;
5. Subaru Telescope, National Astronomical Observatory of Japan, 650 North A'ohoku Place, Hilo, HI 96720, USA;
6. Department of Astronomy, Graduate School of Science, The University of Tokyo, 731 Hongo, Bunkyo-ku,
Tokyo 113-0033, Japan;
7. East Asian Observatory, 660 North A'ohoku Place, Hilo, Hawaii 96720, USA;
8. Department of Astronomy, Faculty of Mathematics and Natural Sciences, Institut Teknologi Bandung, Jl. Ganesha 10, Bandung 40132, Indonesia;
9. Center for Computational Science, University of Tsukuba, Tsukuba, Ibaraki 305-8577, Japan;
10. Department of Earth and Planetary Sciences, Faculty of Sciences, Kyushu University, Fukuoka 812-8581, Japan;
11. ALMA Project Office, National Astronomical Observatory of Japan, Osawa 2-21-1, Mitaka, Tokyo 181-8588, Japan;
12. Nobeyama Radio Observatory, National Astronomical Observatory of Japan, Minamimaki, Minamisaku, Nagano 384-1805, Japan;
13. Department of Earth and Space Science, Osaka University, Machikaneyama-cho, Toyonaka-shi, Osaka 560-0043, Japan;
14. National Astronomical Observatory of Japan, Osawa, 2-21-1, Mitaka, Tokyo 181-8588, Japan
}

\keywords{star formation}
\vskip 0.5cm 
]

\begin{abstract}
We report our current SMA and ALMA studies of disk
and planet formation around protostars.
We have revealed that $r \gtrsim$100 AU scale disks in Keplerian rotation
are ubiquitous around Class I sources. These Class I Keplerian disks are often embedded
in rotating and infalling protostellar envelopes.
The infalling speeds of the protostellar envelopes
are typically $\sim$ 3-times smaller than the free-fall velocities, and the rotational profiles
follow the $r^{-1}$ profile, that is, rotation with the conserved specific angular momentum.
Our latest high-resolution ($\sim$0$\farcs$5) ALMA studies, as well as the other studies in
the literature, have unveiled that $r \sim$100-AU scale
Keplerian disks are also present in several Class 0 protostars,
while in the other Class 0 sources the inferred upper limits of the Keplerian disks
are very small ($r \lessim$20 AU).
Our recent data analyses of the ALMA long baseline
data of the Class I-II source HL Tau have revealed gaps in molecular gas as well as in dust in the surrounding disk, suggesting the presence of sub-Jovian planets in the disk.
These results imply that disk and planet formation should
be completed in the protostellar stage.
\end{abstract}

\section{Introduction}

Two of the most intriguing astrophysical questions are when and how
planets form. A classical picture is that planetary systems form out of
protoplanetary disks around T-Tauri stars. Recent theoretical studies,
however, predict that planets can form when
the central stars are still in the protostellar (Class 0-I) phase,
due to the gravitational instabilities of massive disks surrounding the protostars
\cite{vor11,tsu13,tsu15}. Observational investigations of disk and planet formation
around protostars, however, have been difficult as
they are deeply embedded within protostellar envelopes. The advent of new (sub)millimeter
interferometers, SMA, and ALMA, has now enabled us to perform high-resolution and high-sensitivity
observations of protostellar sources, to identify disks
around protostars, and to study the formation mechanisms of disks and then planets.

We have been conducting systematic SMA and ALMA observations of protostars.
Our group effort aims to unveil the time scale of planet formation, a long-standing
but unresolved astrophysical question.
Our perspective is that disk and planet formation will be initiated
and completed during the protostellar stage and before the T-Tauri stage,
as is supported by our observational results described below.

%

\section{Class I}
\begin{figure*}
\vskip -0.5cm
\centering
$\begin{array}{cc}
\includegraphics[angle=0,height=7.cm]{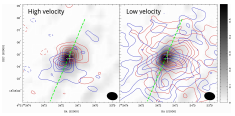} 
\end{array}$
\caption{\it 
Distributions of the high- and low-velocity blueshifted (blue contours) and
redshifted (red) CS (7--6) emission in the Class I protostar L1551 IRS 5
obtained with the SMA and ASTE, superposed on the 343 GHz continuum image (gray scale).
This figure is taken from Figure 5 by Chou et al. (2014).
}
\label{fig:irs5cs}
\end{figure*}

\ \ \ \ In Figure \ref{fig:irs5cs}, SMA+ASTE images of the submillimeter CS (7--6) emission
toward the Class I protostellar binary L1551 IRS 5 (contours), superposed on the SMA image
of the 0.9-mm dust continuum emission (gray), are shown \cite{cho14}.
In the high-velocity region, the blueshifted and redshifted CS emission
trace the southeastern and northwestern parts of the dust-continuum emission,
and exhibit a velocity gradient along the major axis of the dust-continuum emission.
In contrast, in the low-velocity region the blueshifted and redshifted CS emission
are extended beyond the continuum emission, and they
overlap significantly. The centroid location of the blueshifted emission is
slightly shifted to the south, and that of the redshifted emission to the west
and northwest.

\begin{figure*}
\vskip -0.5cm
\centering
$\begin{array}{cc}
\includegraphics[angle=0,height=12.5cm]{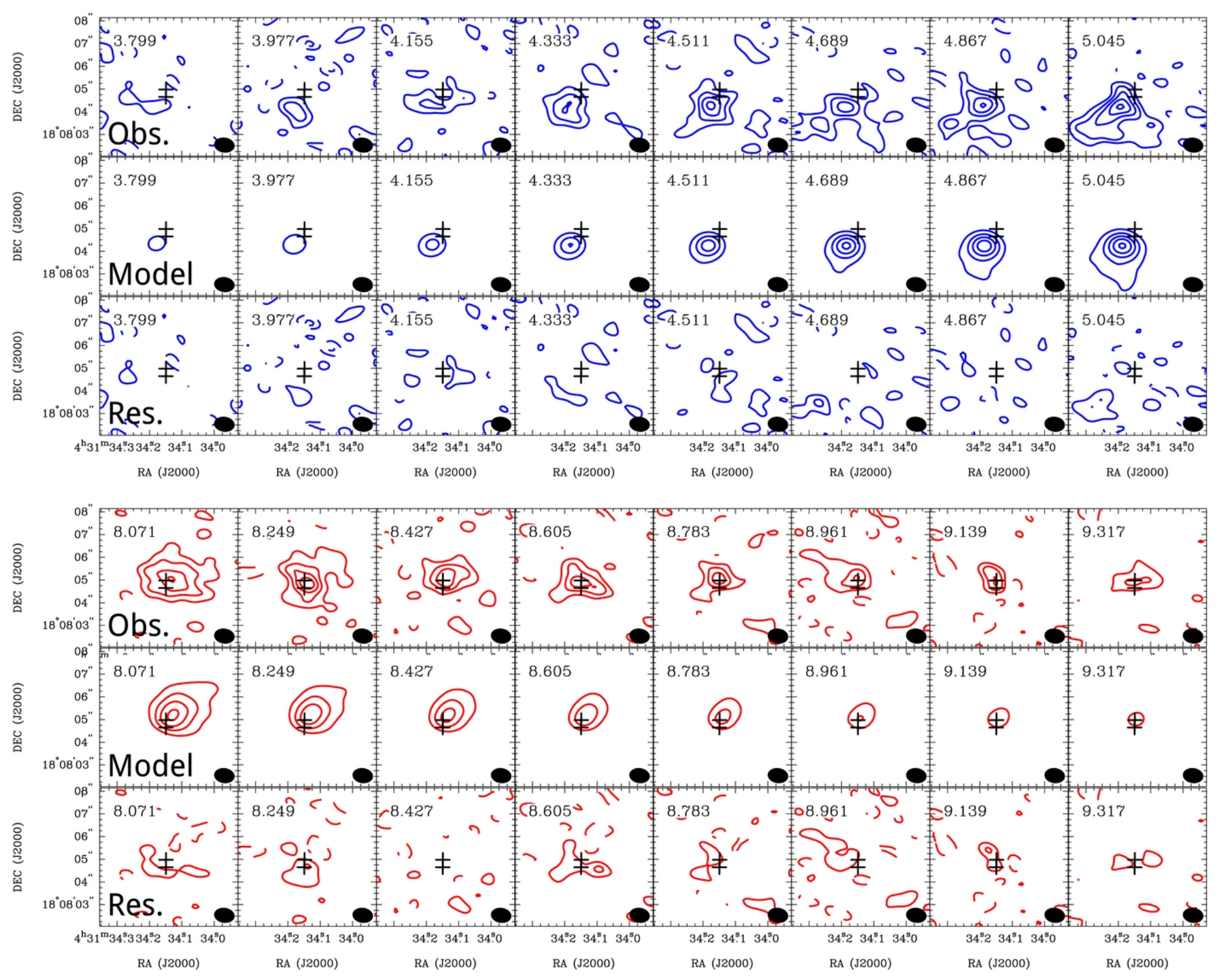} 
\end{array}$
\caption{\it
Best-fit results of the $\chi^2$ model fitting
of the geometrically-thin Keplerian disk to the SMA+ASTE
CS (7--6) velocity channel maps in L1551 IRS 5 at the
highly blueshifted (blue contours) and redshifted (red) velocities.
Upper, middle, and lower panels show the observed, model, and
the residual velocity channel maps,
respectively, where the best-fit parameters are $M_{\star} = 0.5~M_{\odot}$,
the disk inclination angle i = -60$^\circ$, and the disk position angle
$\theta$ = -33$^\circ$, respectively.
This figure is taken from Figure 7 by Chou et al. (2014).
}
\label{fig:csch}
\end{figure*}

To interpret the observed velocity structures in the submillimeter CS emission,
we made a simple geometrically-thin disk + envelope model. Figure \ref{fig:csch}
shows the results of the $\chi^2$ model fitting to the high-velocity blueshifted
and redshifted CS emission. Our model fitting demonstrates that the observed high-velocity
CS emission is well reproduced by a geometrically-thin Keplerian-disk model with
the central stellar mass $M_{\star}$ = 0.5 $M_{\odot}$ and the outermost radius
$R_{kep}$ = 64 AU. On the other hand, the spatial-velocity distribution of the low-velocity
CS emission can be modeled with a rotating and infalling envelope. The envelope rotation
follows the $v_{rot} \propto r^{-1}$ law, that is, rotation with the conserved specific
angular momentum, and the angular momentum of the infalling envelope connects smoothly
to that at the outermost radius of the central Keplerian disk.
The infalling velocity in the envelope is a factor $\sim$3 slower than the free-fall velocity estimated
from the central protostellar mass, which is derived from the central Keplerian rotation.

Our SMA and ALMA observations have also identified $r \sim$100-300 AU scale
Keplerian disks around the Class I protostars L1551 NE \cite{tak12,tak13,tak14,tak15,tak16},
TMC-1A \cite{aso15}, and L1489 IRS \cite{yen13,yen14}. The surrounding envelopes exhibit
an $r^{-1}$ rotational profiles in TMC-1A and L1489 IRS,
while no clear rotation is seen in the envelope around L1551 NE.
All these protostellar envelopes show infalling motion $\sim$3-times slower than
the corresponding free-fall except for L1489 IRS, where the envelope consists of
free-fall gas streams onto the central Keplerian disk.
Table 1 summarizes these results.

\section{Class 0}
\begin{figure*}
\centering
$\begin{array}{cc}
\includegraphics[angle=0,height=5.0cm]{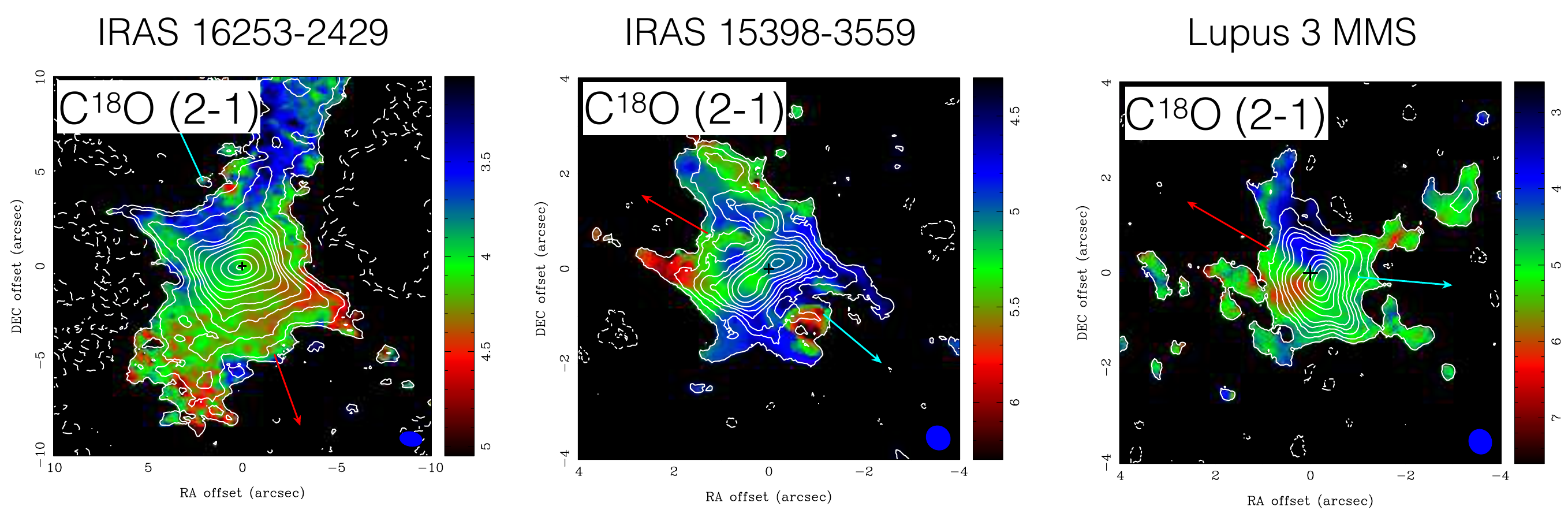} 
\end{array}$
\caption{\it 
Moment 0 (white contours) and 1 maps (colors)
of the C$^{18}$O (2--1) emission toward
three Class 0 protostars observed with ALMA.
Crosses denote the positions of the protostars, and
blue and red arrows the directions of the blueshifted
and redshifted outflows, respectively.
This figure is taken from Yen et al. (2017).
}
\label{fig:c0alma}
\end{figure*}

\ \ \ \ The above results, as well as the other results in the
literature \cite{lom08,bri13,lin14}, demonstrate that
Keplerian disks are ubiquitous around Class I sources.
To clarify how early such Keplerian disks are formed around protostars, we have also
conducted high-resolution ($\lessim 0\farcs5$) ALMA observations of Class 0 protostars
\cite{yen16b}.
Figure \ref{fig:c0alma} shows moment 0 (contours) and 1 (colours) maps of the C$^{18}$O (2--1)
emission toward three Class 0 protostars in the southern sky, IRAS 16253-2429,
IRAS 15398-3559, and Lupus 3 MMS. While these protostars are all Class 0 protostars
with low bolometric temperatures ($T_{bol} \lessim 60~K$), the velocity structures of
their circumstellar materials differ significantly. In IRAS 16253-2429 the protostellar envelope
exhibits a velocity gradient along the direction of the outflow, while in IRAS 15398-3559 velocity
gradients both along and across the outflow axis are seen. In Lupus 3 MMS the northwestern
and southeastern parts show blueshifted and redshifted emission, respectively, and the velocity
gradient is approximately perpendicular to the outflow axis.

\begin{figure*}
\vskip -0.5cm
\centering
$\begin{array}{cc}
\includegraphics[angle=0,height=11.0cm]{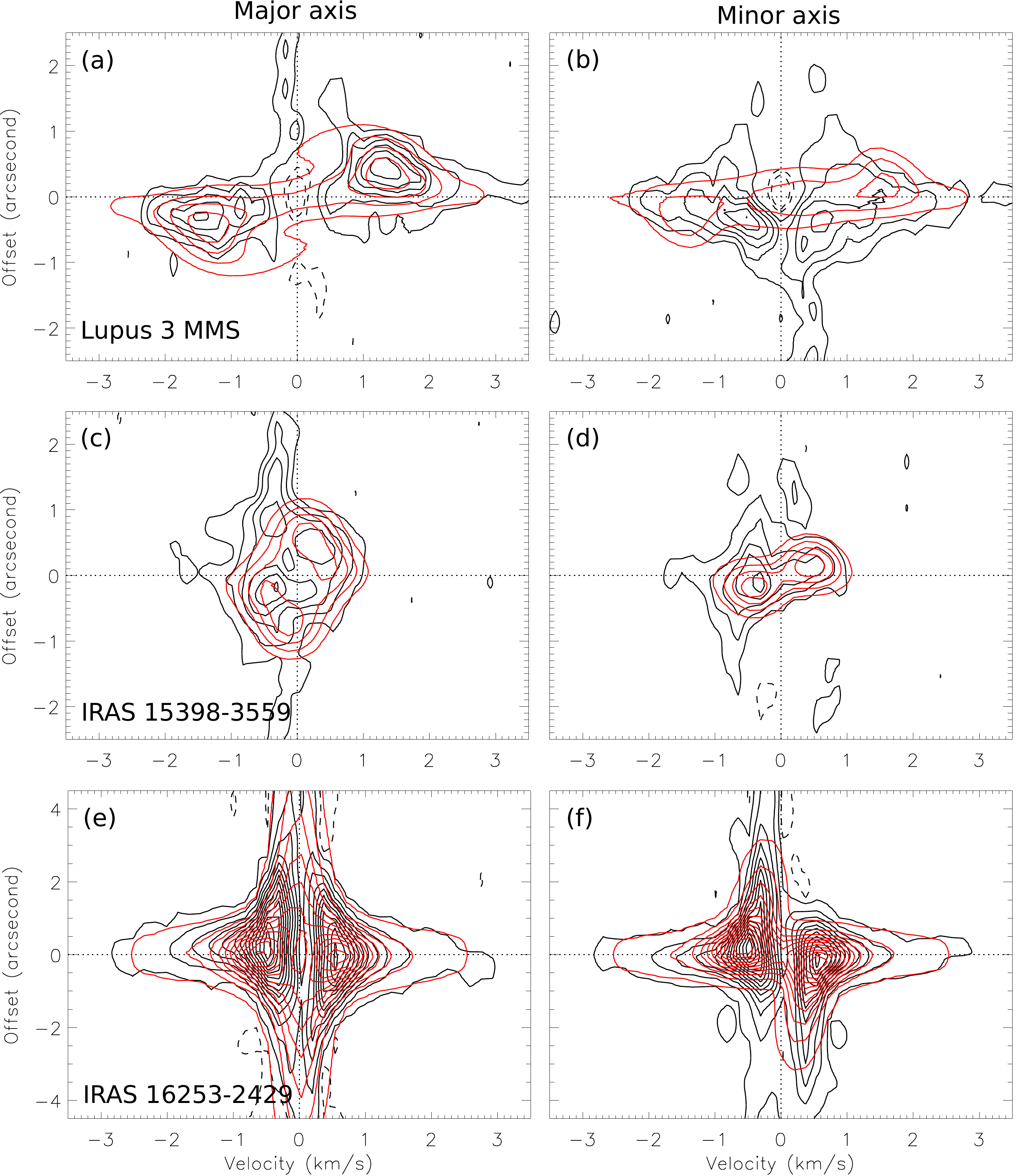} 
\end{array}$
\caption{\it Observed (black contours) and model (red)
Position-Velocity (P-V) diagrams of the C$^{18}$O (2--1) emission toward
the three Class 0 protostars along the major and minor axes of the
protostellar envelopes. This figure is taken from Yen et al. (2017).
}
\label{fig:pv}
\end{figure*}

Figure \ref{fig:pv} shows the observed (black contours) and model (red)
Position - Velocity (P-V) diagrams of the three Class 0 protostars
in the C$^{18}$O emission along (left panel) and across the outflow axes (right).
In Lupus 3 MMS, along the major
axis the blueshifted and redshifted emission are separated and
located to the northwestern and southeastern parts, respectively. It is also clear
that the higher-velocity emission resides closer to the central protostar. Along
the minor axis no clear velocity gradient is seen, while the velocity width becomes
wider near the protostellar position. Our model (red contours) shows that the
observed spatial and velocity features are reproduced by a Keplerian disk
with $M_{\star}$ = 0.3 $M_{\odot}$ and $R_{kep} =$130 AU.
In IRAS 15398-3559, velocity gradients
are seen both along and across the outflow axes. Our model reveals that these features
can be interpreted by an infalling envelope with $r^{-1}$ rotation.
Toward IRAS 16253-2429, on the other hand, only the velocity gradient along the outflow axis
can be identified, suggesting infall without a clear rotation.

\begin{figure}
\vskip 0.5cm
\centering
$\begin{array}{cc}
\includegraphics[angle=0,width=8.0cm]{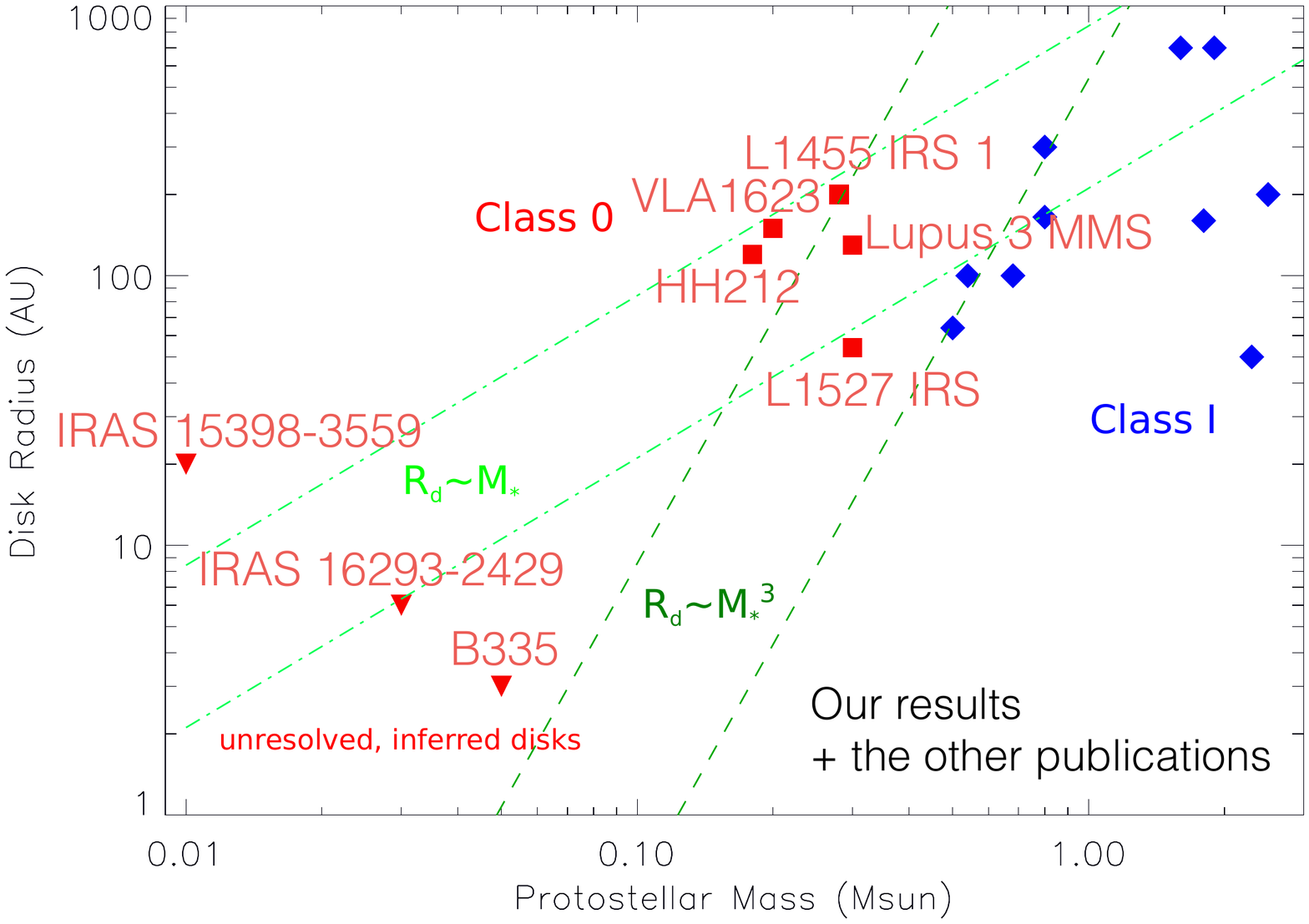}
\end{array}$
\caption{\it Plots of the protostellar masses versus the outermost radii
of the Keplerian disks for a sample of Class 0 and I protostars.
Red squares and blue diamonds present Class 0 and I protostars
associated with the resolved Keplerian disks.
Red downward triangles denote
the Class 0 protostars without direct detections of the Keplerian disks.
Their inferred protostellar masses and disk radii are upper limits.
Dark and light green lines denote the scaling relations between protostellar masses
and disk radii in the collapse models by Terebey et al. (1984) and Basu (1998), respectively.
This figure is taken from Yen et al. (2017).
}
\label{fig:plot}
\end{figure}

Figure \ref{fig:plot} shows our compilation of the central stellar masses versus
the radii of the Keplerian disks ($M_{\star} - R_{kep}$) for a sample of Class 0 and
I protostars from our previous observations and from the literature. Red and blue
symbols denote the results of the Class 0 and I sources, respectively.
A number of Class I and several Class 0 protostars possess large ($r \gtrsim$50 AU)
Keplerian disks, and the protostellar masses of the Class 0 sources
are systematically lower than those of the Class I sources. Besides,
there is another group of Class 0 protostars (IRAS 15398-3559, IRAS 16293-2429, and B335)
with very small upper limits of the Keplerian radii ($\lessim$20 AU) and
the protostellar masses ($\lessim$0.05 $M_{\odot}$).
These results imply
that the formation of $r \sim$100 AU scale Keplerian disks should be completed
by the late Class 0 stage ($\sim10^{5}~yr$).

\begin{table*}[htb]
\scriptsize
\begin{center}
\caption{Summary of our SMA and ALMA Observations of Protostellar Sources}
  \begin{tabular}{|l|ccp{3cm}cp{3.2cm}|} \hline
    Protostar & $T_{bol}$ (K) & $M_{\star} (M_{\odot})$ & Envelope &$R_{kep}$ (AU)
    &Our Publications \\ \hline
    \multicolumn{6}{|c|}{Class 0} \\ \hline
    B335            &31  &0.05 &Infall, Slow Rotation &$<$3 &Yen et al. 2010; 2011; 2013; 2015a,b \\
    IRAS 16293-2429 &36  &0.03 &Infall, Slow Rotation &$<$6 &Yen et al. 2017 \\
    Lupus 3 MMS     &39  &0.3  &Keplerian Disk only   &130  &Yen et al. 2017 \\
    L1527 IRS       &59  &0.3  &Slow Infall, $r^{-1}$ Rotation &54 &Yen et al. 2013; 2015a; Ohashi et al. 2014\\
    IRAS 15398-3559 &61  &0.01 &Infall, $r^{-1}$ Rotation   &$<$20 &Yen et al. 2017 \\ \hline
    \multicolumn{6}{|c|}{Class I} \\ \hline
    L1551 NE        &91  &0.8  &Slow Infall, Slow Rotation &300 &Takakuwa et al. 2012; 2013; 2014; 2015; 2016 \\
    L1551 IRS 5     &92  &0.5  &Slow Infall, $r^{-1}$ Rotation &64 &Takakuwa et al. 2004; Chou et al. 2014 \\
    TMC-1A          &172 &0.64 &Slow Infall, $r^{-1}$ Rotation &100 &Yen et al. 2013; Aso et al. 2015 \\
    L1489 IRS       &238 &1.6  &Freefall, $r^{-1}$ Rotation &300 &Yen et al. 2013; 2014 \\ \hline
  \end{tabular}
\end{center}
\end{table*}

\section{HL Tau}

As discussed above, disk formation around protostellar sources is likely to be completed
by the beginning of the Class I phase. Thus, planet formation
may be initiated from as early as the Class I stage. A famous Class I-II source
as a candidate of planet formation is HL Tau \cite{dip15,kan15}, where ALMA partnership et al. (2015) have identified
seven ringlike gaps in the disk as seen in the dust-continuum emission.
The gaps in the dust-continuum emission, however, can be created by spatial variations
of dust properties and they do not necessarily reflect the ``real" gaps of material
\cite{zha15,oku16}. To reveal whether the dust gaps are real material gaps or not
and to discuss whether planet formation is indeed ongoing in the disk,
it is critical to observe the distribution of the 100 times more abundant molecular gas.

\begin{figure}
\vskip -0.5cm
\centering
$\begin{array}{cc}
\includegraphics[angle=0,width=7.5cm]{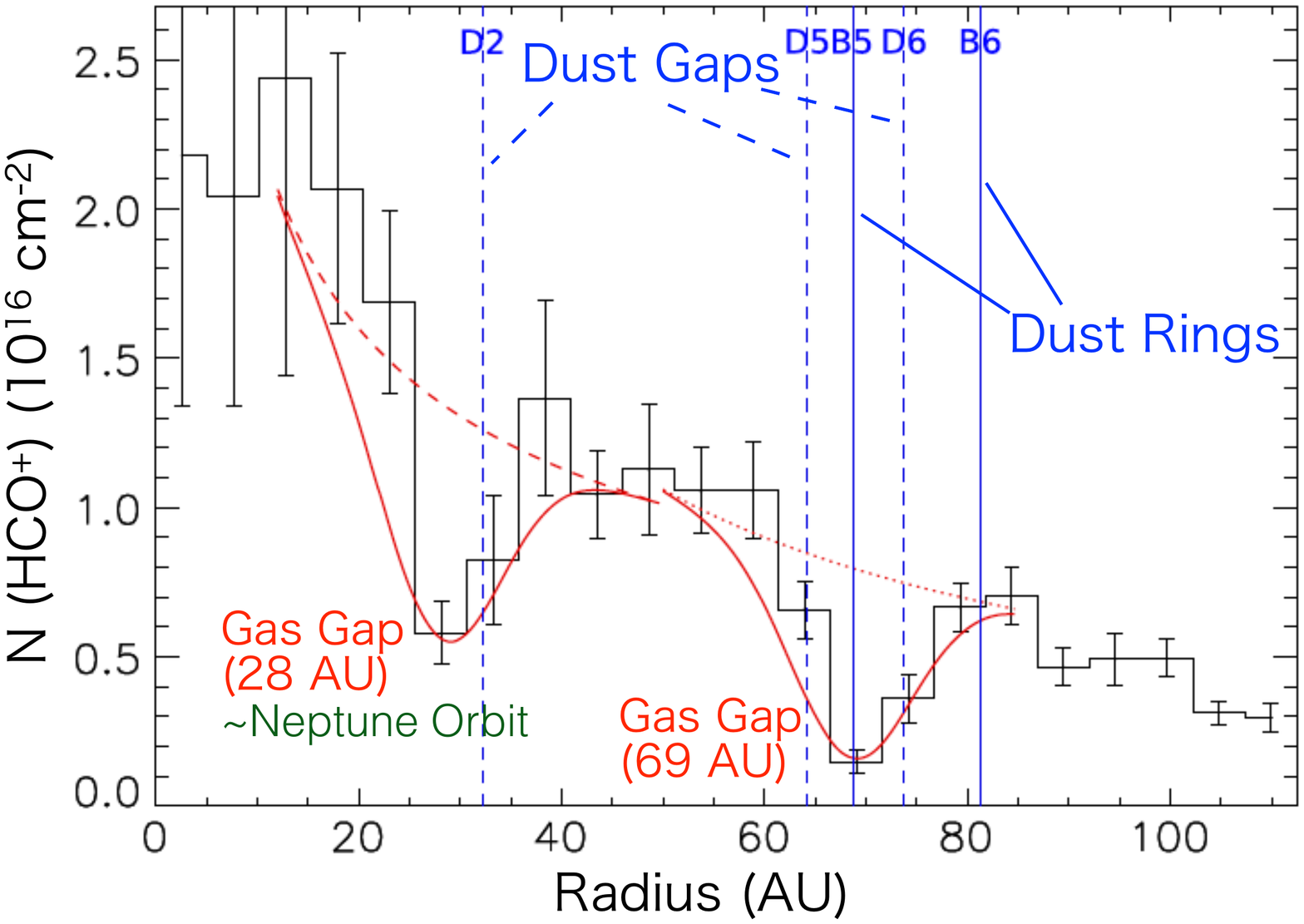} 
\end{array}$
\caption{\it Radial profile of the HCO$^+$ column density in
the protoplanetary disk around HL Tau. Histograms with the error
bars denote the observed values.
Red dashed and dotted lines present the fitted power-law profiles
outside the inner and outer gaps, respectively,
and red solid lines delineate the observed gaps.
This figure is taken from Yen et al. (2016a).
}
\label{fig:hltau}
\end{figure}

Signal-to-noise ratios of molecular-line images are generally much lower
than those of dust-continuum images, because molecular-line images cannot have
wide bandwidths. The lower signal-to-noise ratios result in limited spatial resolutions
of the molecular-line images.
To overcome these problems and to unveil the corresponding
gaps of molecular lines in the disk around HL Tau, we made a high-resolution ($\sim$0$\farcs$1$\times$0$\farcs$05)
HCO$^{+}$ (1--0) image cube from the same ALMA dataset as that of the dust gaps, and
performed annular averaging of the  HCO$^{+}$ image cube \cite{yen16a}.
While the annular averaging
loses the azimuthal information, it can produce a high-resolution and high-sensitivity
radial intensity profile, as shown in Figure \ref{fig:hltau}. The derived radial profile
of HCO$^{+}$ clearly exhibits two gaps at radii of $\sim$28 AU ($c.f.,$ orbital radius of Neptune = 30 AU)
and $\sim$69 AU, which correspond to the locations of the dust gaps of D2 and D6.
The FWHM widths of the inner and outer HCO$^{+}$ gaps are both estimated to be
$\sim$14 AU, and the depths are estimated to be a factor $\sim$2.4 and $\sim$5.0,
respectively.
The presence of the two gas gaps
in the inner two dust gaps implies that these two gaps are real gaps of material.
The most interesting interpretation is that these gaps trace orbits
of planet bodies which are sweeping material along their orbits,
although secular gravitational instability can also create such gaps \cite{inu14}.
From the observed widths and depths of the inner and outer gaps, the masses of the putative planet
bodies are estimated to be $\sim$0.8$M_J$ and $\sim$2.1$M_J$, respectively \cite{kan15}.

These results imply that planet formation can be initiated from as early as
the Class I stage. Our group has an approved follow-up ALMA project to directly
measure the gas gaps without the annular averaging in the higher-$J$ HCO$^{+}$ transitions,
and ALMA and ACA data of the envelope surrounding the HL Tau disk.
With these data we will continue to study the details of the planet-forming system.

\section{Summary}

We have been conducting SMA and ALMA studies of disk- and planet formation
around protostars. We consider that disk and planet formation should be completed
within the protostellar phase. If our working hypothesis could be proven observationally,
this should impact our understanding of planet formation significantly.
So far, we have obtained the following results, which are indeed supporting
our hypothesis.

\begin{itemize}
\item[1.] $r \gtrsim$100 AU-scale Keplerian disks are ubiquitous around Class I protostars.
Such disks are also seen in several Class 0 protostars, while in the other Class 0 protostars
the upper limits of the Keplerian radii are very small ($r \lessim$20 AU). Thus, there
is an apparent bimodality of Keplerian radii among the Class 0 samples. These results
indicate rapid growth of Keplerian disks during the short Class 0 stage ($\sim 10^{5}~yr$).
 
\item[2.]
Those protostellar sources, both with and without large Keplerian disks, are often embedded in
infalling envelopes. In many sources the infalling envelopes also exhibit rotating motions,
and the rotational profiles
follow the $r^{-1}$ profile, that is, rotation with conserved specific angular momentum.
The angular momentum of the infalling envelope connects smoothly
to that at the outermost radius of the central Keplerian disk. On the other hand,
the infalling velocity is a factor $\sim$3 slower than the free-fall velocity estimated
from the central protostellar mass, which is derived from the central Keplerian rotation.

\item[3.]
We have re-analyzed the ALMA long baseline data of HL Tau in the HCO$^{+}$ (1--0) emission.
With the annular averaging, we have made a high-resolution ($\sim$10 AU) radial profile of the HCO$^{+}$
column density, and have found two regions deficient of molecular gas (``gaps") at
radii of $\sim$28 AU and $\sim$69 AU, which are consistent with the locations of the
dust gaps. These results imply that these two gaps of molecular gas and dust are real
gaps of material.
The FWHM widths of the inner and outer HCO$^{+}$ gaps are both estimated to be
$\sim$14 AU, and the depths are estimated to be a factor $\sim$2.4 and $\sim$5.0.
Assuming that these gaps are created by planetary bodies, the inferred planet masses
are $\sim$0.8 $M_J$ and $\sim$2.1 $M_J$ at radii of 28 and 69 AU, respectively.


\end{itemize}


\section*{Acknowledgments}
S.T. and P.M.K. acknowledge grants from the Ministry of Science and Technology (MOST) of Taiwan
(MOST 102-2119-M-001-012-MY3) and (MOST 103-2119-M-001-009), respectively.
S.T. is grateful to a grant JSPS KAKENHI Grant Number JP16H07086,
in support of this work.
This paper makes use of the following ALMA data:
ADS/JAO.ALMA\#2013.1.00879.S and ADS/JAO.ALMA\#2011.0.00015.SV.
ALMA is a partnership of ESO (representing its member states), NSF (USA) and NINS (Japan),
together with NRC (Canada) and NSC and ASIAA (Taiwan), in cooperation with
the Republic of Chile. The Joint ALMA Observatory is operated by ESO, AUI/NRAO and NAOJ.


\end{document}